\documentstyle{elsart}

\input epsf

\begin{document}
\begin{frontmatter}
\title{Quantifying dynamics of the financial correlations}
\author[IFJ,KFA]{S. Dro\.zd\.z},
\author[IFJ] {J. Kwapie\'n},
\author[KFA]{F. Gr\"ummer},
\author[WLB]{F. Ruf},
\author[KFA,ADELA]{J. Speth}

\address[IFJ]{Institute of Nuclear Physics, Radzikowskiego 152,
31--342 Krak\'ow, Poland}
\address[KFA]{Institute f\"ur Kernphysik, Forschungszentrum J\"ulich,
D--52425 J\"ulich, Germany}
\address[WLB]{West LB International S.A., 32--34 bd
Grande--Duchesse Charlotte, L--2014 Luxembourg}
\address[ADELA]{Centre for the Subatomic Structure of Matter,
University of Adelaide, SA 5005, Australia}

\begin{abstract}
A novel application of the correlation matrix formalism to study dynamics of the
financial evolution is presented. This formalism allows to quantify 
the memory effects as well as some potential repeatable intradaily structures 
in the financial time-series. The present study is based on the high-frequency
Deutsche Aktienindex (DAX) data over the time-period between November 1997
and December 1999 and demonstrates a power of the method.
In this way two significant new aspects of the DAX evolution are
identified: (i) the memory effects turn out to be sizably shorter than what the
standard autocorrelation function analysis seems to indicate and (ii)
there exist short term repeatable structures in fluctuations that are governed
by a distinct dynamics. The former of these results may provide an argument
in favour of the market efficiency while the later one may indicate origin 
of the difficulty in reaching a Gaussian limit, expected from the central 
limit theorem, in the distribution of returns on longer time-horizons.
\end{abstract}
\end{frontmatter}


\section{Introduction}

In global terms the financial correlations can be classified 
as correlations in space and correlations in time, though of course
they are somewhat interrelated. The first category so far
studied involves for instance the correlations among the companies 
comprised by a single stock market~\cite{Lalo,Pler,Dro1}, among a group
of subjects~\cite{Mant,Kull} and even between the different and geographically
remote stock markets~\cite{Dro2}. From practical perspective this type 
of correlations relates to the theory of optimal portfolios~\cite{Mark,Elto} 
and risk management. In the context of the stock market an important result 
of this study is that majority of eigenvalues in the spectrum of the
correlation matrix agree very well with the universal predictions of
random matrix theory~\cite{Wign,Meht}.
Locations of some of the eigenvalues differ however from these predictions 
and thus, similarly as in other physical systems~\cite{Dro3,Dro4}, 
identify certain system-specific, non-random properties such as collectivity.

Studying explicit correlations in time is at least as important because 
it is this type of correlations that directly reflects a nature of the 
financial dynamics. In general the character of those correlations is however
less understood and several related issues still remain puzzling.     
It is for instance commonly accepted~\cite{Camp} that the autocorrelation 
function of the financial time-series drops down to zero within few minutes
which reflects a time-horizon of the market inefficiency. At the same time
the correlations in volatility remain positive for many weeks.
On short time-scales the return distributions are definitely not Levy
stable~\cite{Stan} but it turns out difficult to detect their convergence
to a Gaussian on longer time-scales as expected from the central limit theorem.
In this connection it is also appropriate to mention a still poorly
understood phenomenon of log-periodicity~\cite{Sor1,Dro5} 
which seems to reflect existence of some very specific correlations 
on all time-scales.
As a contribution towards resolving this sort of difficulties below we 
propose to use the concept of the correlation matrix such that it focuses
entirely on the time-correlations and their potential existence can 
parallelly be detected on various time-scales. Utility of such a procedure 
is illustrated on an example of high-frequency (15sec) recordings 
of the Deutsche Aktienindex (DAX). The entries of the correlation matrix 
are then constructed from the time-series of returns representing the 
consecutive trading days. As a result such entries are labelled by the pairs
of different days. Several striking observations are also made.          

\section{Methodology}

The relevant correlation matrix is defined as follows. To each element in
a certain sequence $N$ of relatively long consecutive time-intervals of
equal length $T'$ labelled with $\alpha$ one uniquely assigns a time
series $x_{\alpha}(t_i)$, where $t_i$ $(i=1,...,T')$ is to be understood
as discrete time counted from the beginning for each $\alpha$. In the
present case $\alpha$ is going to label the consecutive trading days and
$t_i$ the trading time during the day. Similar methodology has already
been successfully applied~\cite{Kwap} to extract the repeatable structures
in the brain sensory response.

If, as here, $x_{\alpha}(t_i)$ represents a price time-series than it is
natural to define the returns $G_{\alpha}(t_i)$ time-series as
\begin{equation}
G_{\alpha}(t_i) = \ln x_{\alpha}(t_i+\tau) - \ln x_{\alpha}(t_i)
\simeq {x_{\alpha}(t_i+\tau) - x_{\alpha}(t_i) \over x_{\alpha}(t_i)}
\ ,
\label{eq:ret}
\end{equation}
where $\tau$ is the time lag. The normalized returns, with the average
value subtracted and its variance normalized to unity, are defined by
\begin{equation}
g_{\alpha}(t_i) = {G_{\alpha}(t_i) - \langle G_{\alpha}(t_i) \rangle_t
\over v^2} \ , \quad
v = \sigma(G_{\alpha}) = \sqrt{\langle G_{\alpha}^2(t) \rangle_t -
\langle G_{\alpha}(t) \rangle_t^2} \ ,
\label{eq:normret}
\end{equation}
where $v$ is volatility of $G_{\alpha}(t)$ and $\langle\ldots\rangle_t$
denotes averaging over time. One thus obtains $N$ time series
$g_{\alpha}(t_i)$ of length $T$ ($T$=$T'$-1), {\it i.e.} an $N \times T$
matrix $\bf M$. Then, the correlation matrix is defined as ${\bf C} = (1 /
T) \ {\bf M} {\bf M}^{\bf T}$. By diagonalizing $\bf C$ 
\begin{equation}
{\bf C} {\bf v}^k = \lambda_k {\bf v}^k,
\label{eq:eig}
\end{equation}
one obtains the eigenvalues $\lambda_k$ $(k=1,...,N)$ and the corresponding
eigenvectors ${\bf v}^k = \{ v^k_{\alpha} \}$.

In the limiting case of entirely random correlations the density of
eigenvalues $\rho_C(\lambda)$ defined as 
\begin{equation}
\rho_C(\lambda) = {1 \over N} {{dn(\lambda)} \over {d \lambda}},
\label{eq:rho}
\end{equation}
where $n(\lambda)$ is the number of eigenvalues of $\bf C$ less than $\lambda$,
is known analytically~\cite{Edel}, and reads
\begin{eqnarray}
\hspace{2.5cm}
\rho_C(\lambda) = {Q \over {2 \pi \sigma^2}} 
{\sqrt{ (\lambda_{max} - \lambda) (\lambda - \lambda_{min}} \over
{\lambda}}, \\
\centering
\lambda^{max}_{min} = \sigma^2 (1 + 1/Q \pm 2 \sqrt{1/Q}),
\nonumber
\label{eq:rho1}
\end{eqnarray}
with $\lambda_{min} \le \lambda \le \lambda_{max}$, $Q=T/N \ge 1$, and
where $\sigma^2$ is equal to the variance of the time series (unity in our
case).

\section {Results}

Our exploratory study along the above indicated line is based on the DAX 
recordings with the frequency of 15 sec during the period between 
November 28th, 1997 and December 30th, 1999. By taking the DAX intraday
variation between the trading time 9:03 and 17:10 which corresponds to
$T=1948$ one then obtains $N=517$ complete and equivalent time series
representing different trading days during this calendar period. (Several
days with incomplete recordings have been rejected.) Using this set of
data we construct the $517 \times 517$ matrix $\bf C$.

\begin{figure}
\epsfxsize 6cm
\hspace{0.5cm}
\epsffile{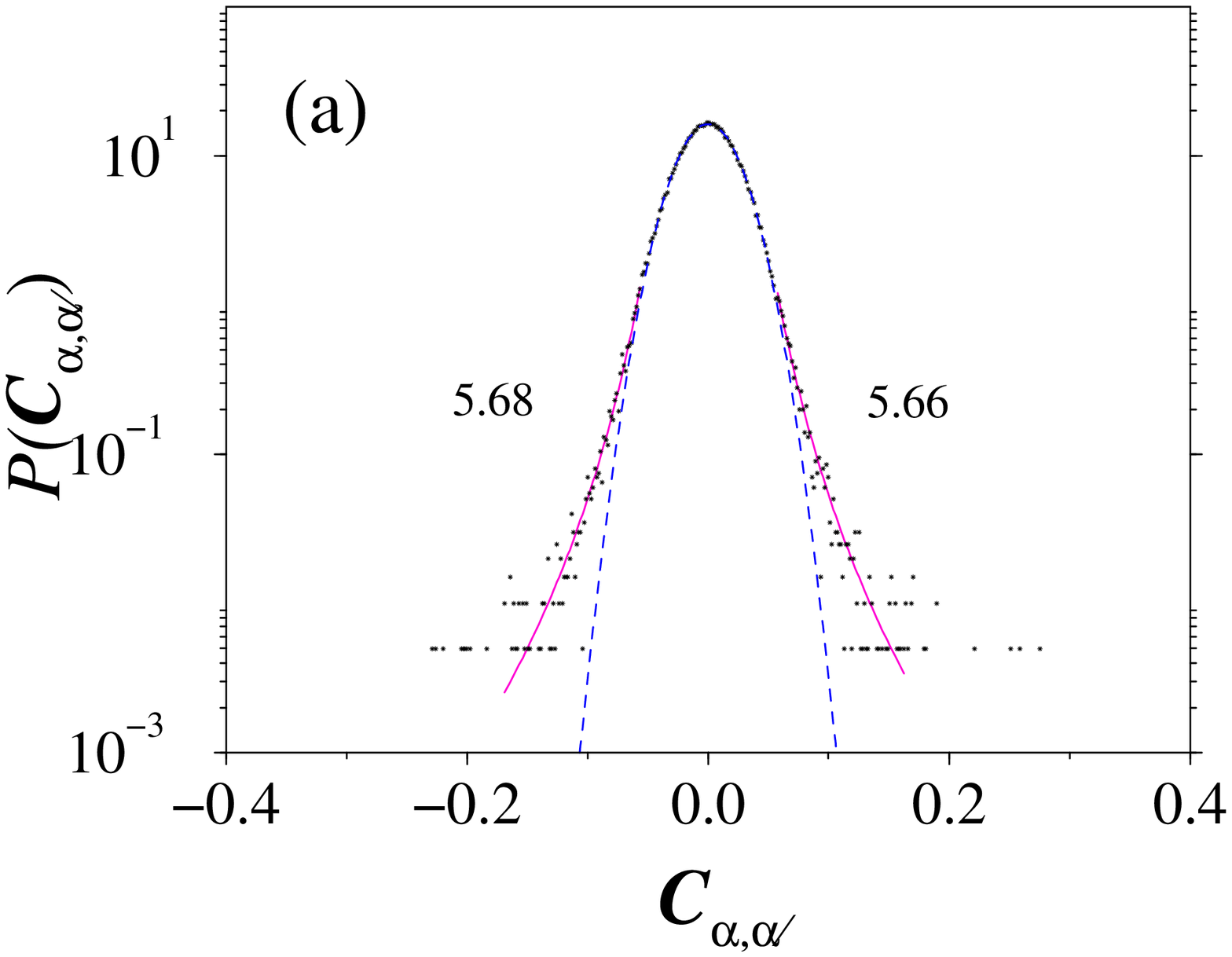}
\hspace{0.5cm}
\epsfxsize 6cm
\epsffile{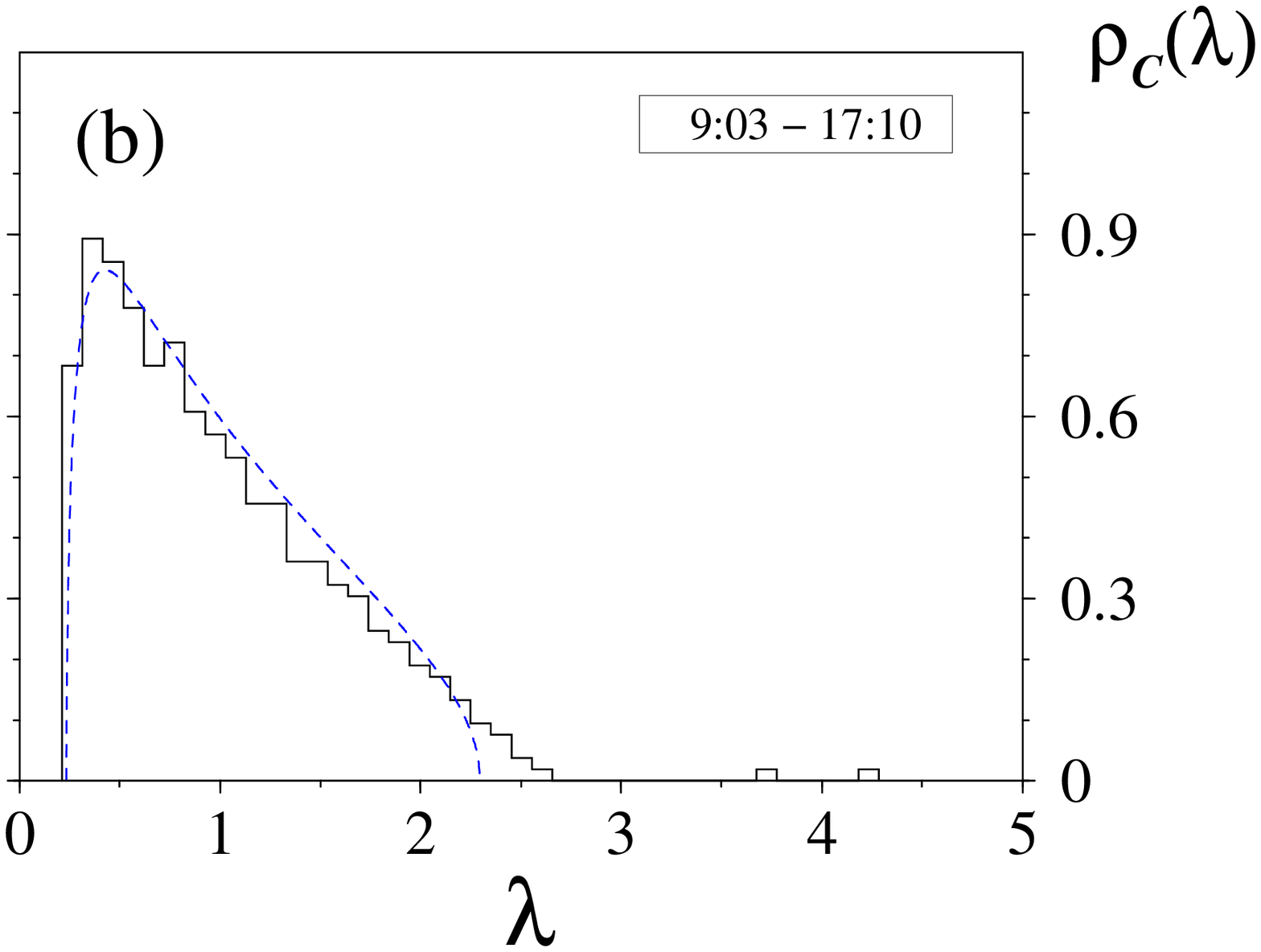}
\caption{$517 \times 517$ correlation matrix $\bf C$ calculated from the
15 sec frequency DAX variation during 9:03 -- 17:10 intraday trading time.
$N=517$ is the number of trading days (labelled by $\alpha$) qualified
for this study from the calendar period November 28th, 1997 -- December
30th, 1999. The dashed line represents the best fit in terms of a Gaussian
while the solid line indicates a power law fit to the tails of this
distribution. The numbers reflect the corresponding scaling indices.
(b) The probability density (histogram) of the eigenvalues of $\bf C$
and the corresponding null hypothesis (dashed line) formulated in terms of
eq.~(\ref{eq:rho}).}
\label{fig:fig1}
\end{figure}

A characteristics of primary interest is the structure of eigenspectrum of
$\bf C$. The structure of eigenspectrum of a matrix is expected to be
related~\cite{Dro3,Dro4} to the distribution of its elements. The
distribution of such elements of $\bf C$ corresponding to the above
specified procedure and the data set is shown in the left panel of
Fig.~\ref{fig:fig1} Clearly, this distribution is symmetric with respect
to zero, a Gaussian like (dashed line) on the level of small matrix
elements, but sizably thicker than a Gaussian on the level of large matrix
elements. As indicated in the figure, in the latter case, a power law with
the index of about 5.6 (consistent with the distribution of returns) 
provides a reasonable representation. This is far beyond the Levy stable
regime but points to the two ensembles of random matrices, the
Gaussian~\cite{Wign,Meht,Edel} and the Levy distributed~\cite{Cize,Burd}, 
as interesting limits for formulating the null hypotheses.   

The resulting probability density of eigenvalues, shown in the right panel
of Fig.~\ref{fig:fig1}, displays a somewhat unexpected structure. There
exist two outliers significantly above $\lambda_{max}$ (for $Q= 1948/517$,
$\lambda_{max} \approx 2.3$) which seem to indicate that the dynamics
under consideration is not a pure noise, but instead that certain
time specific repeatable structures in the intraday trading do take place.  
However, it is even more astonishing that the bulk of the spectrum 
agrees very well with the bounds prescribed by purely random correlations.
This indicates that the neighbouring recordings in our time series of 15 sec
DAX returns share essentially no common information. Our previous
experience~\cite{Kwap} teaches us that the probability density of
eigenvalues of the correlation matrix is a sensitive measure of such effects.
Typically, a common information shared by neighbouring events results in
an effective number ($T_{eff}$) of time points such that $T_{eff}$ is
significantly smaller than $T$. This immediately affects both
$\lambda_{max}$ and $\lambda_{min}$. In the present case we see basically
no such effect, particularly on $\lambda_{min}(\approx 0.23)$ side.
A whole nonrandomness can be here associated with the two largest
eigenvalues.

\begin{figure}
\centering
\epsfxsize=10cm
\mbox{\epsfbox{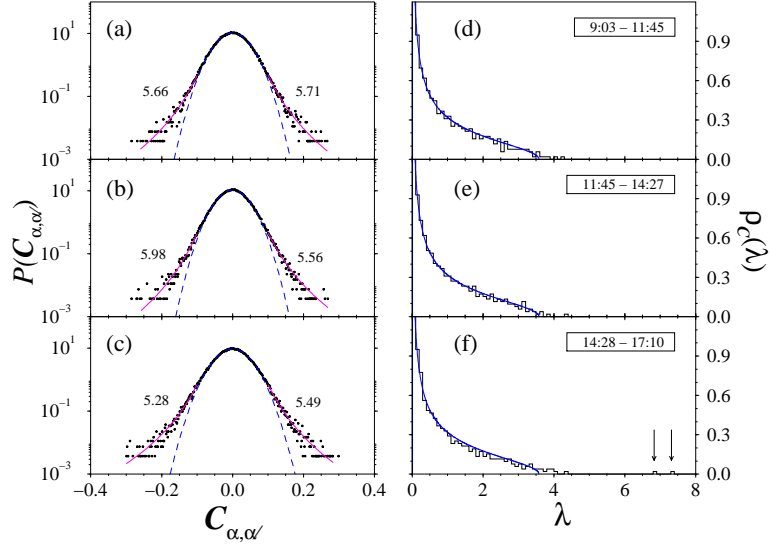}}
\caption{(Left column) Distribution of matrix elements $C_{\alpha,\alpha'}$ 
of the $517 \times 517$ correlation matrix $\bf C$ calculated from 
the 15 sec frequency DAX variation during the intraday trading time 
9:03--11:45 (a), 11:45--14:28 (b) and 14:28-17:10 (c). Similarly as in
Fig.~\ref{fig:fig1} $N=517$ is the number of trading days (labelled by
$\alpha$) qualified for this study from the calendar period November
28th, 1997 -- December 30th, 1999. The solid lines indicate the power law
fits to the tails of these distributions, while the dashed one corresponds
to a Gaussian best fit. The numbers reflect the corresponding scaling
indices. (Right column) The probability densities (histograms) of the
eigenvalues of $\bf C$ (d), (e) and (f) corresponding to the three (a),
(b) and (c) cases, respectively. Their null hypotheses formulated in terms
of eq.~(\ref{eq:rho}) are indicated by the continuous lines.} 
\label{fig:fig2}
\end{figure}

In order to identify the character of eigenstates  associated with
these eigenvalues we split the whole daily time interval 
considered into three equal subintervals. For each of them $T=649$
(thus $\lambda_{min} \approx 0.01$ and $\lambda_{max} \approx 3.58$). 
The resulting distributions of matrix elements of $\bf C$ and the corresponding
probability densities of eigenvalues are shown in Fig.~\ref{fig:fig2}.
It appears easy to see that separation of the two largest eigenvalues 
originates from the last subinterval. Consistently, it is also this
subinterval which produces the thickest tails in the distribution of large
matrix elements.

\begin{figure}
\centering
\hspace{-0.5cm}
\epsfxsize=12.0cm
\epsfysize=8.0cm
\mbox{\epsfbox{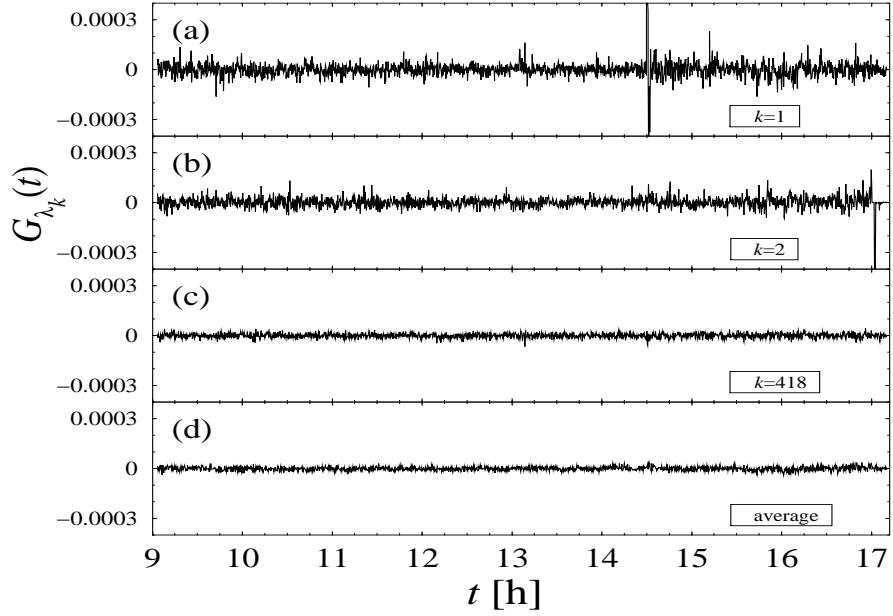}}
\caption{The superposed time series of unnormalized returns calculated
according to eq.~(\ref{eq:gs}) for $k=1$ (a), $k=2$ (b) and $k=418$ (c),
and of the simple average (d) of the original returns as functions of the
intraday trading time.}
\label{fig:fig3}
\end{figure}

An even more efficient way to visualize the differences among the 
eigenvectors is to look at the superposed time series of returns
\begin{equation}
G_{{\lambda}_k}(t_i) = \sum_{\alpha=1}^N {\mathrm sign}(v^k_{\alpha}) 
\vert v^k_{\alpha} \vert^2 G_{\alpha}(t_i).
\label{eq:gs}
\end{equation}
In this definition $\vert v^k_{\alpha} \vert^2$ is used 
instead of $v^k_{\alpha}$ for the reason of preserving normalization and
the sign of $v^k_{\alpha}$ to account for a possible coherence of the
original signals. A collection of such superposed time series of returns 
for $k=1$, 2 and 418, and of the simple average of the original returns
is shown in Fig.~\ref{fig:fig3}. For a statistical value of $k$, as it is
illustrated by an example of $k=418$ ((c) in Fig~\ref{fig:fig3}),
$g_{{\lambda}_k}(t_i)$ basically does not differ from the simple
average (panel (d) in Fig.~\ref{fig:fig3}). The first two differ however
significantly and indicate the existence of repeatable structures at the
well defined instants of time through many days. Quite unexpectedly, the
most collective signal, associated with the largest eigenvalue ($k=1$
shown in (a) of Fig.~\ref{fig:fig3}), reflects the strongest synchronous
DAX activity precisely at around 14:30 (probably in response to the
North-American financial news release exactly at this time) and not at the
time when the Wall Street opens, nor even just before closing in
Frankfurt. This last period of an enhanced activity also carries some
signatures of synchrony but is entirely comprised by the second
eigenstate.
   
\begin{figure}
\centering
\epsfxsize=8.0cm
\mbox{\epsfbox{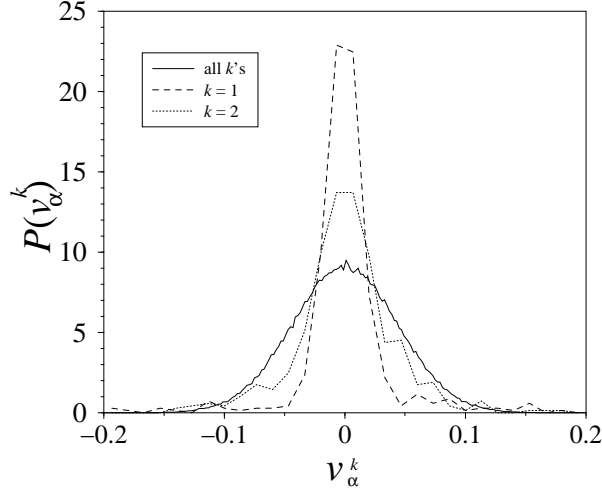}}
\caption{Distribution of eigenvector components $v_{\alpha}^k$ for the
two most collective eigenstates ($k=1$ and $k=2$) together with the
corresponding distribution of components for all the eigenstates.}
\label{fig:fig4} \end{figure}

Another characteristics which carries the relevant information is the
probability distribution of the eigenvector components $v^k_{\alpha}$.
Such a probability distribution evaluated from all $k$'s is illustrated 
by the solid line in Fig.~\ref{fig:fig4}. It globally is well represented
by a Gaussian. Significantly different are the distributions of eigenvector
components for our two outliers; they are concentrated more at around zero
but, at the same time, the tails of the distributions are thicker.
This effect is more pronounced for $k=1$ than for $k=2$ which indicates
that fewer days ($\alpha$'s) contribute to the strong signal seen at 14:30
for $k=1$ than just after 17:00 for $k=2$.

\begin{figure}
\centering
\epsfxsize=12.0cm
\mbox{\epsfbox{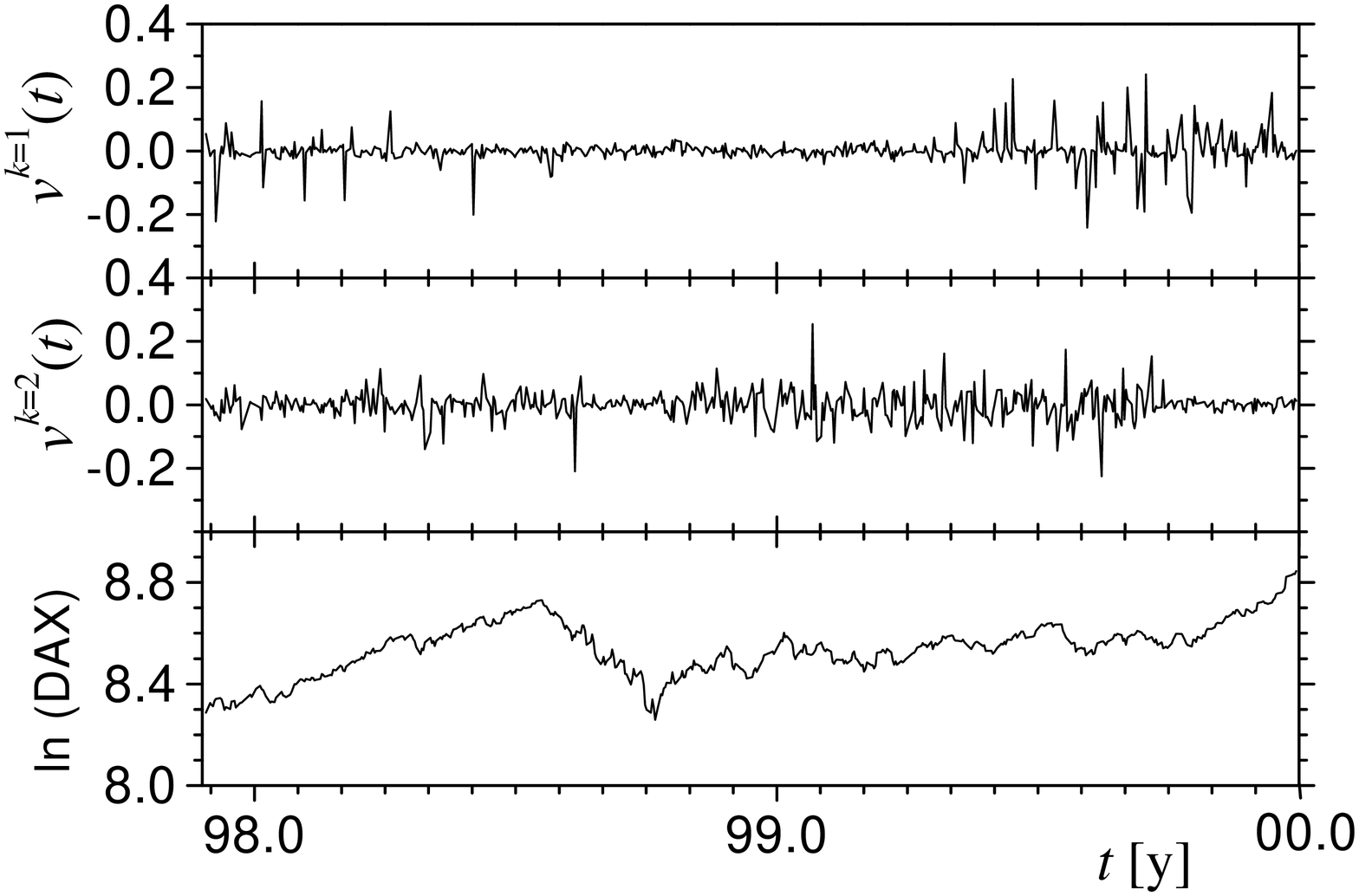}}
\caption{Eigenvector components $v_{\alpha}^k$ for $k=1$ (upper) and
$k=2$ (middle) as a function of trading days from the entire period
November 28th, 1997 - December 30th, 1999. For a comparison, the logarithm
of DAX time variation is presented in the bottom panel.}
\label{fig:fig5}  
\end{figure}

This issue is quantified in more detail in Fig.~\ref{fig:fig5} which shows 
how these two eigenvector components are distributed in magnitude over
the trading days $(\alpha$'s) incorporated in the present study.  
For $k=1$ such components assume large values indeed and even seem to
be developing certain periodicity, which in the initial period considered
here corresponds to about 20 trading days (about one calendar month).
Interestingly, it is this period which corresponds to our classic example of 
"Imprints of log-periodic self-similarity in the stock market"~\cite{Dro5}.   
Perhaps the observed, equidistant in time, synchronous bursts of activity
constitute one of the elements introducing a characteristic time-scale
responsible for discrete scale invariance~\cite{Sor2}, an element which in
natural way may contribute to the appearance of the log-periodic
oscillations.

\begin{figure}
\centering
\epsfxsize=8.0cm
\mbox{\epsfbox{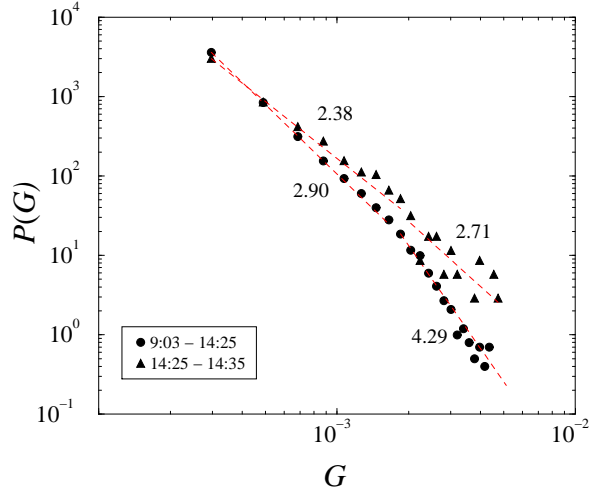}}
\caption{Tails of the return distribution through all 517 trading days
for the periods 9:03-14:25 and 14:25-14:35. The corresponding linear best 
fits indicate (see also the corresponding numbers) two different regimes
(dashed lines).}
\label{fig:fig6}
\end{figure}

Finally, we find it appropriate to notice that the return probability
density distributions associated with the above identified synchronous
bursts of activity seem to be governed by a different law than what more
global statistical analyses~\cite{MaSt} document. Fig.~\ref{fig:fig6}
shows the tails of the return $G_{\alpha}(t_i)$ distribution (triangles)
for the time period 14:25--14:35 through all the DAX 517 trading days
considered here. The statistics is of course poor but still it quite
convincingly indicates much thicker tails as compared to the periods of
'normal' activity, represented in this figure (circles) by the period
9:03--14:25. In the latter case the pdf parameters are about consistent
with those cited in the literature~\cite{MaSt}, while in the short time
interval at around 14:30 we seem to be still in the L\'evy stable regime.         

\section {Summary}

The present study sheds a somewhat new light on the issue of time-correlations
in the dynamics of financial evolution. It shows for instance that the
consecutive returns carry essentially no common information even when
probed with the frequency of 15 sec. Such a conclusion sounds somewhat
in contrast to a common belief, based on autocorrelation function analysis,
that the market inefficiency time-horizon extends over few minutes. There
is however perhaps no contradiction between our conclusion and the 
time-lag dependence of the autocorrelation function. This function remains
positive for about few minutes indeed, but already after 15 sec it drops down 
by at least one order of magnitude. On the other hand, however, at the
well defined short periods of time during the intraday trading, there
exist clearly identifiable synchronous repeatable bursts of activity (for
DAX at 14:30) whose related return probability density functional develops
significantly larger values on the level of rare events than during
periods of the normal activity. This, together with the fact that such
events are strongly correlated in time, may constitute a principal reason
for an observed extremely slow convergence to a Gaussian of the global
return distribution on long-time scales. It would now be very interesting
to perform similar study for the other stock market indices as well.


\begin{thebibliography}{9}

\bibitem{Lalo} L.~Laloux, P.~Cizeau, J-.P~Bouchaud, M.~Potters,
Phys.~Rev.~Lett. {\bf 83} (1999) 1467.

\bibitem{Pler} V.~Plerou, P.~Gopikrishnan, B.~Rosenow, L.A.N.~Amaral,
H.E.~Stanley, Phys.~Rev.~Lett. {\bf 83} (1999) 1471.

\bibitem{Dro1} S.~Dro\.zd\.z, F.~Gr\"ummer, A.Z.~G\'orski, F.~Ruf,
J.~Speth, Physica A {\bf 287} (2000) 440

\bibitem{Mant} R.N.~Mantegna, Eur.~Phys.~J.~B {\bf 11} (1999) 193

\bibitem{Kull} L.~Kullmann, J.~Kert\'esz, R.N.~Mantegna,
Physica A {\bf 287} (2000) 412

\bibitem{Dro2} S.~Dro\.zd\.z, F.~Gr\"ummer, F.~Ruf, J.~Speth,
{\it Towards identifying the world stock market cross-correlations: 
DAX versus Dow Jones}, arXiv:cond-mat/0011488, Physica A in press

\bibitem{Mark} H.~Markowitz, Portfolio Selection:
Efficient Diversification of Investments, J. Wiley and Sons, New York, 1959

\bibitem{Elto} E.J.~Elton and M.J.~Gruber, Modern Portfolio Theory and
Investment Analysis, J.~Wiley and Sons, New York, 1995

\bibitem{Wign} E.P.~Wigner, Ann. Math. {\bf 53} (1951) 36.

\bibitem{Meht} M.L.~Mehta, Random Matrices, Academic Press, Boston, 1999

\bibitem{Dro3} S.~Dro\.zd\.z, S.~Nishizaki, J.~Speth, M.~W\'ojcik,
Phys. Rev. E {\bf 57} (1998) 4016.

\bibitem{Dro4} S.~Dro\.zd\.z, M.~W\'ojcik,
{\it On the origin of order from random two-body interactions},
arXiv:nucl-th/0007045

\bibitem{Camp} J.Y.~Campbell, A.W.~Lo, A.~Craig MacKinley, 
The Econometrics of Financial Markets, Princeton University Press, 
Princeton, NJ, 1997

\bibitem{Stan} H.E.~Stanley, P.~Gopikrishnan,~V. Plerou, L.A.N.~Amaral,
Physica A {\bf 287} (2000) 339

\bibitem{Sor1} D.~Sornette, A.~Johansen, J.-P.~Bouchaud, J.~Phys. I France
{\bf 6} (1996) 167; \\
J.A.~Feigenbaum, P.G.O.~Freund, Int.~J.~Mod.~Phys.~B {\bf 10} (1996) 3737

\bibitem{Dro5} S.~Dro\.zd\.z, F.~Ruf, J.~Speth and M.~W\'ojcik,
Eur. Phys. J. {\bf B10} (1999) 589.

\bibitem{Kwap} J.~Kwapie\'n, S.~Dro\.zd\.z, A.A.~Ioannides,
Phys. Rev. {\bf E62} (2000) 5557

\bibitem{Edel} A.~Edelman, SIAM J.~Matrix Anal.~Appl. {\bf 9} (1988) 543;
\\ 
A.M.~Sengupta, P.P.~Mitra, Phys.~Rev.~E {\bf 60} (1999) 3389

\bibitem{Cize} P.~Cizeau, J.-P.~Bouchaud, Phys.~Rev.~E {\bf 57} (1994)
1810

\bibitem{Burd} Z.~Burda, R.A.~Janik, J.~Jurkiewicz, M.A.~Nowak, G.~Papp,
I.~Zahed, {\it Free Random Levy Matrices}, arXiv:cond-mat/0011451 

\bibitem{Sor2} D.~Sornette, Phys.~Rep. {\bf 297} (1998) 239

\bibitem{MaSt} R.N.~Mantegna, H.~Eugene Stanley, An Introduction to
Econophysics: Correlations and Complexity in Finance, University Press,
Cambridge, 2000

\end{thebibliography}
\end{document}